\documentclass{llncs}
\usepackage{makeidx}  
\usepackage{graphicx}
\usepackage{subfigure}
\usepackage{amssymb,amsmath,bm}
\usepackage{mathrsfs}
\usepackage{booktabs}
\usepackage{multirow}
\usepackage{float}
\usepackage{textcomp}
\usepackage{color}
\usepackage{marvosym}
\usepackage{todonotes}
\begin{document}
	\title{Deformable Registration of Brain MR Images via a Hybrid Loss}
	
	\author{Luyi Han\inst{1} \and
			Haoran Dou\inst{2} \and
			Yunzhi Huang\inst{3}\textsuperscript{(\Letter)} \and
			Pew-Thian Yap\inst{4}\textsuperscript{(\Letter)}}
	\institute{
		Department of Radiology and Nuclear Medicine, Radboud University Medical Center, Geert Grooteplein 10, 6525 GA, Nijmegen, The Netherlands\\ \and
		Centre for Computational Imaging and Simulation Technologies in Biomedicine (CISTIB), University of Leeds, UK\\ \and
		College of Biomedical Engineering, Sichuan University, Chengdu, China
		\email{yunzhi.huang.scu@gmail.com}\\ \and
		Department of Radiology and Biomedical Research Imaging Center (BRIC), University of North Carolina, Chapel Hill, USA\\
		\email{ptyap@med.unc.edu}}
	
	\maketitle

	\begin{abstract}
		Unsupervised learning strategy is widely adopted by the deformable registration models due to the lack of ground truth of deformation fields. These models typically depend on the intensity-based similarity loss to obtain the learning convergence.  
		Despite the success, such dependence is insufficient. For the deformable registration of mono-modality image, well-aligned two images not only have indistinguishable intensity differences, but also are close in the statistical distribution and the boundary areas. 
		Considering that well-designed loss functions can facilitate a learning model into a desirable convergence, we learn a deformable registration model for T1-weighted MR images by integrating multiple image characteristics via a hybrid loss.
		Our method registers the OASIS dataset with high accuracy while preserving deformation smoothness.
	\end{abstract}

	\section{Introduction}
	Deformable registration estimates dense deformation fields to establish image-to-image correspondence. Conventional methods typically involve time-consuming iterative optimization and experience-dependent parameter tuning. Alternatively, deformations can be learned for fast registration via (1) supervised learning~\cite{rohe2017svf,sokooti2017nonrigid,cao2018deformable,eppenhof2018pulmonary}; (2) weakly-supervised learning~\cite{hu2018weakly}; and (3) unsupervised learning~\cite{balakrishnan2018unsupervised,hoopes2021hypermorph}. 
	
	Supervised learning methods rely on deformations predicted using conventional methods (\textit{e.g.}, SyN~\cite{avants2008Symmetric} or Diffeomorphic Demons~\cite{D.Demons}) and simulations~\cite{cao2018deformable,eppenhof2018pulmonary}. 
	In contrast, weakly-supervised and unsupervised learning methods do not require ground truth deformations.  
	Weakly-supervised learning methods optimize model parameters via supervision using label-level similarity and segmentation maps to align structural boundaries~\cite{hu2018weakly}. 
	Unsupervised learning methods are supervised via intensity-level similarity (\textit{e.g.}, Normalized Cross Correlation (NCC) or Sum of Squared Difference (SSD)). 
	
	Our method combines weakly-supervised and unsupervised learning and learns registration via multiple aspects, including intensity, statistics, label levels. 
	The proposed method ranked fifth on the brain T1w deformable registration task organized by the MICCAI 2021 Learn2Reg challenge\footnote{https://learn2reg.grand-challenge.org}.
	
	\section{Method}
\begin{figure}[t]
		\label{fig:framework}
		\centering
		\includegraphics[width=1.3\textwidth]{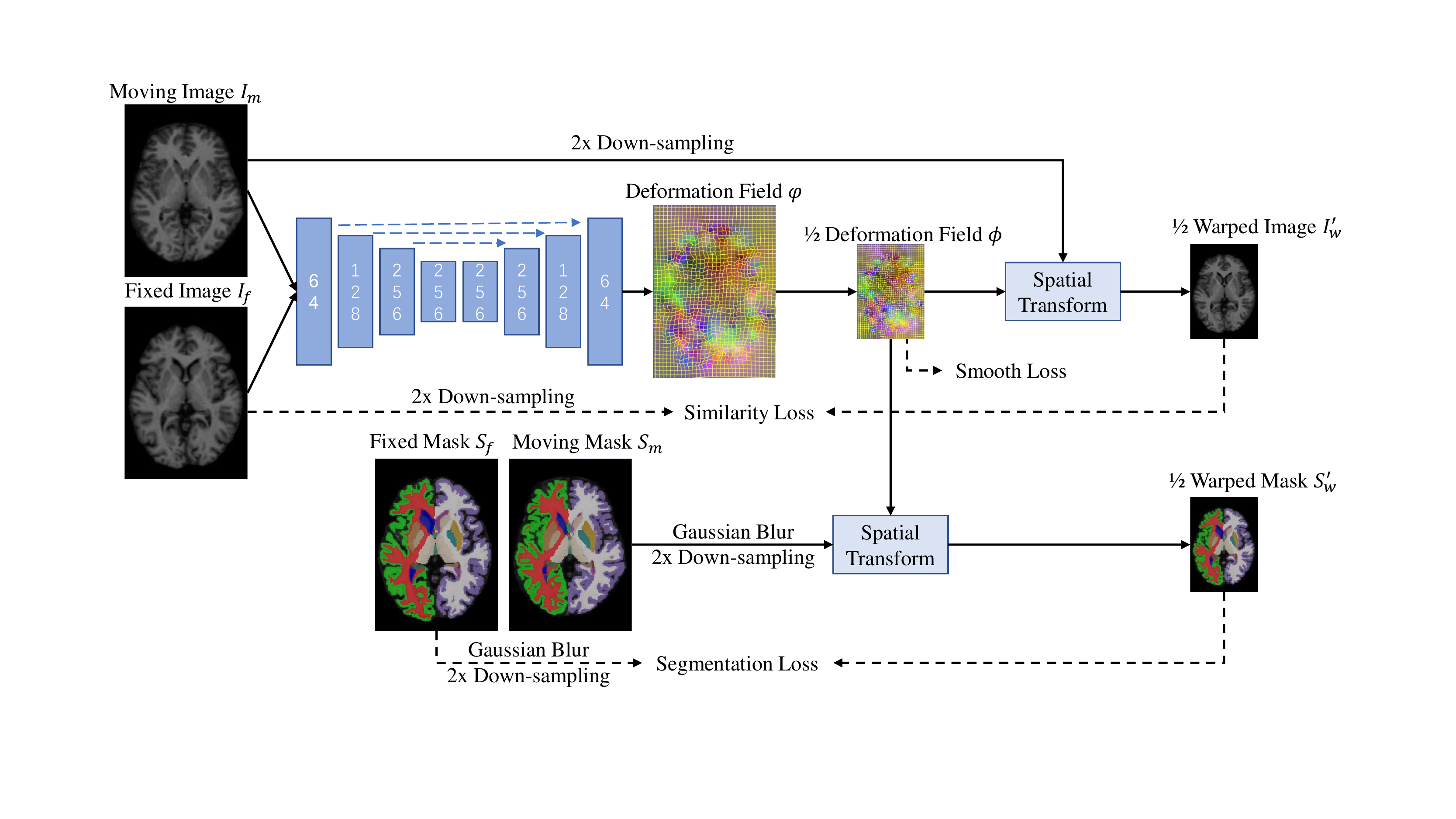}
		\caption{Overview of the proposed registration model. Random selected patches from the inter-subject T1w image pairs input the registration network to output the deformation field. During training, similarity loss, segmentation loss, and smooth loss are bonded together to guide the learning.}
	\end{figure}

	The core of our deformable registration model (Fig.~\ref{fig:framework}) is based on VoxelMorph~\cite{balakrishnan2018unsupervised} with the following modifications: (1) increased feature channels for each layer and (2) deformation field downsampling by a factor of 2. 
	The input is a randomly selected pair of patches from the moving and fixed images. The output is the predicted \textit{x}, \textit{y}, and \textit{z} displacements at half the resolution of the input. 
	
	\subsection{Hybrid Loss}
	Aligned images should be matched at the boundary, intensity, and statistical distribution levels. We employ multi-faceted supervision involving a hybrid loss function to improve the alignment between the moving image $I_{m}$ and the fixed image $I_{f}$.
	
	\subsubsection{Intensity Loss}
	We employ the commonly used SSD to gauge the intensity dissimilarity between $I_{m}$ and $I_{f}$:
	\begin{equation}
		\label{eq:intensity loss}
		\mathcal{L}_{i} = || I_{w}^{'} -I_f^{'} ||^2_{2}
	\end{equation}
	where $I_{w}^{'} = I_m^{'}\circ\phi$ is the half size moving image $I_m^{'}$ warped with predicted displacement field $\phi$. $I_{m}^{'}$ and $I_f^{'}$ are downsampled from the original moving image $I_{m}$ and fixed image  $I_{f}$ by factor 2.

	\subsubsection{Statistic Loss}
	We employ mutual information~\cite{guo2019multi} to improve the joint probability distribution between $I_{w}^{'}$ and $I_{f}^{'}$:
	\begin{equation}
		\label{eq:statistic loss}
		\mathcal{L}_{s} = H(I_{w}^{'}) + H(I_f^{'})-H(I_{w},I_f^{'})
	\end{equation}
	where $H(\cdot)$ refers to the entropy of an image, and $H(\cdot,\cdot)$ is the joint entropy of two images. 
	
	\subsubsection{Boundary Loss}
	We employ the area overlap between the segmentation mask $S_w^{'}$ of $I_{w}^{'}$ and downsampled segmentation mask $S_f^{'}$ of $I_{f}^{'}$ for boundary-level supervision.
	The segmentation maps are encoded in one-hot format and are convoluted with a Gaussian blur kernel with the size of 7 and $\sigma$ of 1.
	We combine $L1$ and $Dice$ for boundary loss:
	\begin{equation}
		\label{eq:boundary loss}
		\mathcal{L}_{b} = \|S_{w}^{'}-S_f^{'}\|_{1}+(1-\frac{2\|S_{w}^{'} \cdot S_f^{'}\|_{1}}{\|S_{w}^{'} + S_f^{'}\|_{1}})
	\end{equation}
	where $S_{w}^{'} = S_m^{'}\circ\phi$ refers to the warped segmentation map. 
 
	\subsubsection{Total Loss}
	In addition to the losses described above, we include a gradient-based regularization term to preserve the topology of the deformation field. The total loss is
	\begin{equation}
	\label{eq:total loss}
		\begin{aligned}
			\mathcal{L} 
			= \mathcal{L}_{i} + \mathcal{L}_{s} + \mathcal{L}_{b} + \lambda\cdot \texttt{Grad}(\phi)
		\end{aligned}
	\end{equation}
	where $\lambda$ balances the dissimilarity term and regularization term and is set empirically to 0.8 via grid research.
	
	\subsection{Dataset and Implementation Details}
	\paragraph{Dataset}	
	Training (414 subjects), validation (20 subjects), and testing (39 subjects) were based on the Open access series of imaging studies (OASIS) dataset~\cite{marcus2007open} curated by the organizers of Learn2Reg MICCAI Challenge 2021~\cite{hering2021learn2reg}~\footnote{https://learn2reg.grand-challenge.org/Datasets/}.	
	OASIS is a cross-sectional MRI data study with a wide range of participants from young, middle aged, nondemented, and demented older adults. 
	Pre-processing (skull-stripping, normalisation, pre-alignment and resampling) was done according to the procedure described in \cite{hoopes2021hypermorph}. 
	Semi-automatic labels with manual corrections of 35 cortical and subcortical brain structures were generated using 3D Slicer.
	
	\paragraph{Implementation Details}
	We implemented our method using Pytorch with NIVIDIA 3090 RTX. We optimized the model with ADAM, with learning rate $1e-6$, a default of 200,000 steps, and batch size of 1. 
	During training, data was augmented by randomly selecting patches of size $128\times128\times128$ from the input volumes. During testing, a deformation field was predicted for an image volume. 13GB and 11GB of GPU memory was consumed during the training and testing stages, respectively.
	
	\section{Experimental Results}
	
	\subsection{Results}
\begin{figure}[!htbp]
		\label{fig:result}
		\centering
		\includegraphics[width=\textwidth]{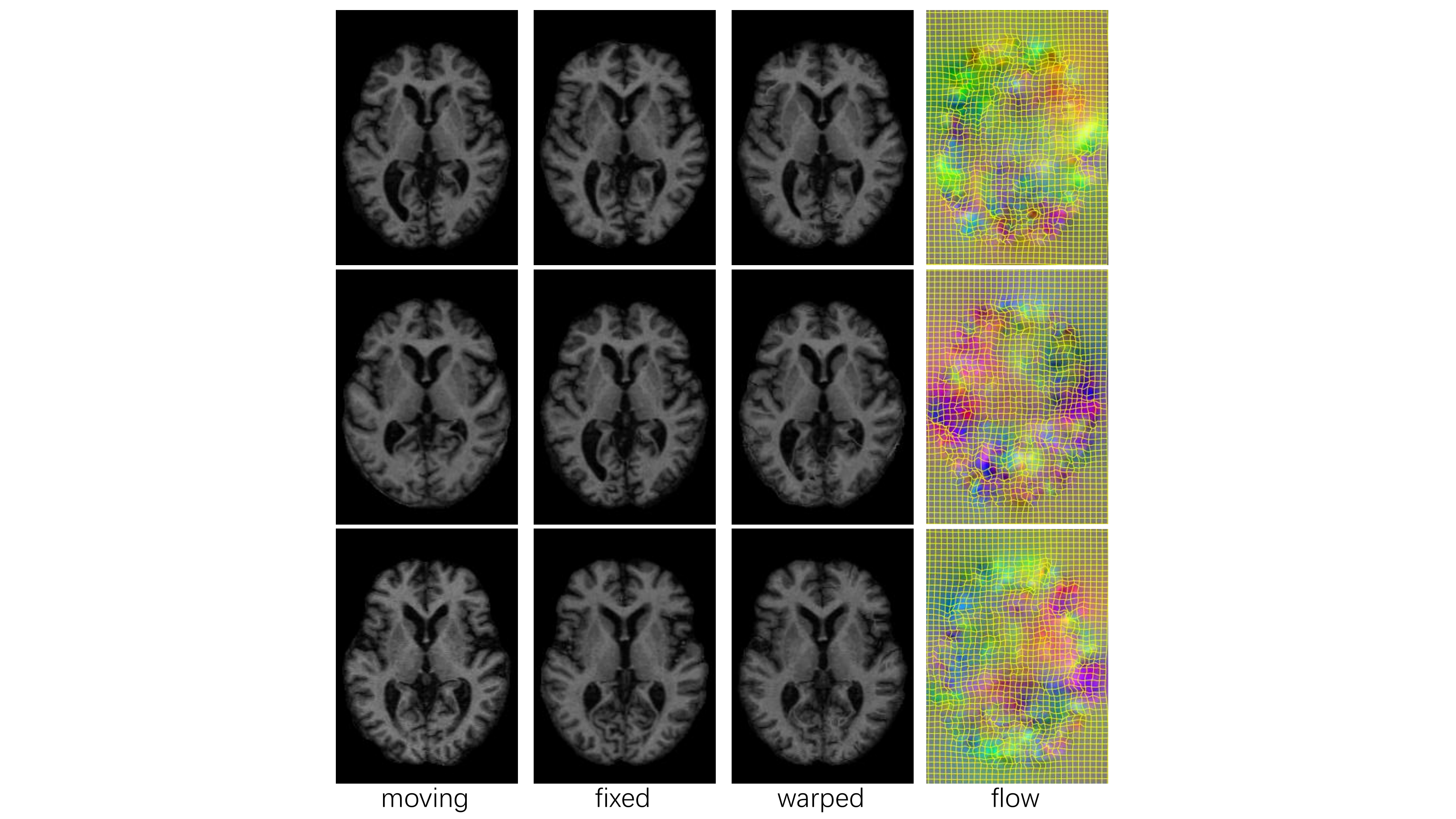}
		\caption{From left to right are the moving image, the fixed image, the warped image, and the predicted deformation field.}
	\end{figure}

	Table~\ref{tab:rst} lists the registration accuracy of different settings on the loss functions.
	Our method achieves for the testing dataset an average Dice score of $80.47\%$ with standard error $1.67\%$ and an average Hausdorff distance of $1.8015\pm0.4325 mm$ over 35 brain ROIs, with SDlogJ of
	$0.0822\pm0.0042$ for full size deformation field $\psi$. 
	Fig.~\ref{fig:result} shows exemplar registration results given by our method. 
	
	\begin{table}
		\centering
		\caption{Ablation study on the validation dataset.}\label{tab:rst}
		\begin{tabular}{|l|l|l|l|l|}
			\hline
			Method & DSC$\uparrow$ & $HD$(mm)$\downarrow$ & SDlogJ$\downarrow$ \\
			\hline
			patch VM-c32 & 0.7978$\pm$0.0230 & 1.9733$\pm$0.4777 & 0.0848$\pm$0.0057 \\
			patch VM-c64 & 0.8040$\pm$0.0209 & 1.9432$\pm$0.4687 & 0.0839$\pm$0.0053 \\
			patch VM-c64+MI+Dice & 0.8117$\pm$0.0214 & 1.8549$\pm$0.4363 & 0.0811$\pm$0.0053 \\
			patch VM-c64+MI+Dice+halved(Proposed) & 0.8395$\pm$0.0142 & 1.6635$\pm$0.3734 & 0.0788$\pm$0.0044 \\
			\hline
		\end{tabular}
	\end{table}
	
	\section{Discussion and Conclusion}
	
		A good deformable registration framework is required to accurately calculate the complex mapping between image pairs.
		The widely used intensity similarity loss is highly dependent on the image quality, requiring the voxels inside each tissue to close at intensity across individual data. Hence, the intensity-driven supervision is efficient yet insufficient. A diverse similarity constraints is then required to enhance the optimization at multiple levels.
		 
		In this Challenge, we adopted a hybrid similarity loss to steer the learning procedure. The intensity-based SSD loss and the statistic-based MI loss steer the registration accuracy at both the local voxels and the global intensity distributions. Meanwhile, the boundary loss improves the match at the boundary regions.
	We showed that registration of T1-weighted images can be registered with high accuracy by enforcing similarity at the intensity, statistic, and boundary levels.

	\section*{Acknowledgment}
	Thanks all the organizers of the MICCAI 2021 Learn2Reg challenge.
	The work was supported in part by the National Natural Science Foundation of China under Grant 6210011424.
	
	%
	\bibliographystyle{splncs}
	\bibliography{refs}
	
\end{document}